\documentclass[11pt]{article}
\usepackage{jheppub}
\usepackage{import}
\usepackage{TemplateMB}
\usepackage{physics}
\def\le{\left}
\def\ri{\right}
\newcommand\ov{\over}
\newcommand\p{\ensuremath{\partial}}
\newcommand{\es}[2] {\begin{equation} \label{#1} \begin{split} #2 \end{split} \end{equation}}

\def\<{\langle}
\def\>{\rangle}
\newcommand\al{{\alpha}}
\newcommand\ep{\varepsilon}

\newcommand\lam{\lambda}
\newcommand\Lam{\Lambda}

\newcommand\ga{{\ensuremath{{\gamma}}}}

\newcommand\de{{\ensuremath{{\delta}}}}
\newcommand\De{{\ensuremath{{\Delta}}}}

\author[a]{Marta Bucca}     
\author[a]{and M\'ark Mezei}                         
 
 \affiliation[a]{Mathematical Institute, University of Oxford, Woodstock Road, Oxford, OX2 6GG, United Kingdom}  

\emailAdd{bucca@maths.ox.ac.uk}       
\emailAdd{mezei@maths.ox.ac.uk}   

\begin{document}

\title{Nonlinear soft mode action for the large-$p$ SYK model}

\abstract{The physics of the SYK model at low temperatures is dominated by a soft mode governed by the Schwarzian action.  In \cite{Maldacena:2016hyu} the linearised action was derived from the soft mode contribution to the four-point function, and physical arguments were presented for its nonlinear completion to the Schwarzian.
In this paper, we give two derivations of the full nonlinear effective action in the large $p$ limit, where $p$ is the number of fermions in the interaction terms of the Hamiltonian. The first derivation uses that the collective field action of the large-$p$ SYK model is Liouville theory with a non-conformal boundary condition that we study in conformal perturbation theory. This derivation can be viewed as an explicit version of the renormalisation group argument for the nonlinear soft mode action in \cite{Kitaev:2017awl}.  The second derivation uses an Ansatz for how the soft mode embeds into the microscopic configuration space of the collective fields. We generalise our results for the large-$p$ SYK chain and obtain a ``Schwarzian chain" effective action for it. These derivations showcase that the large-$p$ SYK model is a rare system, in which there is sufficient control over the microscopic dynamics, so that an effective description can be derived for it without the need for extra assumptions or matching (in the effective field theory sense).}

\maketitle

\section{Introduction and summary}

The Sachdev-Ye-Kitaev  (SYK) model~\cite{KitaevTalks,Sachdev:2015efa,Maldacena:2016hyu} and its generalisations~\cite{Gu:2016oyy,Davison:2016ngz,Fu:2016vas,Gross:2017vhb,Murugan:2017eto,Maldacena:2018lmt,Altland:2019lne,Lian:2019axs,Milekhin:2021sqd,Milekhin:2021cou,Berkooz:2021ehv,Berkooz:2022dfr} provide a rare analytically solvable window into many-body quantum chaos. The computation of out-of-time-ordered correlation functions in these models has led to many new insights into the quantum butterfly effect~\cite{Shenker:2013pqa,Roberts:2014isa,Maldacena:2015waa} and operator growth~\cite{Roberts:2018mnp,Qi:2018bje,Parker:2018yvk}. 
Its low energy description in terms of the Schwarzian effective theory uncovered the Nearly-CFT$_1$ universality class of quantum dynamics, and its holographic dual JT gravity description~\cite{Jensen:2016pah,Maldacena:2016upp}.

In the low temperature limit, the dynamics of the SYK model is dominated by a soft mode that encodes the reparametrisations of time. In quantum mechanics the analog of the infinite dimensional conformal symmetry of two-dimensional CFTs is the reparametrisation symmetry of time: however it cannot be a true symmetry in a system with nontrivial dynamics.\footnote{It is however consistent with $H=0$, which is topological quantum mechanics, the physics of ground states.} The breaking of this symmetry is controlled by the Schwarzian action for the reparametrisations of time with a prefactor that scales linearly with the temperature, which explains the importance of this mode at low temperatures:
\begin{equation}
    \label{eqn:SchwarzianIntro}
    S=-\frac{N\alpha_S}{\beta \mathcal{J}}\int_0^{2\pi} du \, \textrm{Sch}\le[\tan\le(f/ 2\ri),u\ri]\,,
\end{equation}
where $N$ is the number of fermions, $\alpha_S$ is a dimensionless number, $\beta$ is the inverse temperature, $\mathcal{J}$ is the dimensionful coupling strength of the SYK model, $u={2\pi\ov \beta}\tau$, $f(u)$ is the time reparametrisation mode, and   Sch$[h,u]={h'''(u)\ov h'(u)}-\frac32\le({h''(u)\ov h'(u)}\ri)^2$ is the Schwarzian derivative.

The results described above were argued for convincingly in several different ways~\cite{Maldacena:2016hyu,Bagrets:2016cdf,Kitaev:2017awl,Rosenhaus:2018dtp}: we give a quick review of the two most detailed arguments below. The key complication in these derivations is the matching between the UV  and the IR, low temperature behaviour of the collective fields, which can only be done indirectly (and involves the constant $\al_S$ that can only be determined numerically). In the large $p$ limit of the SYK model the UV to IR connection is under much better control. In this paper we capitalise on this fact to give two explicit analytical derivations of the nonlinear Schwarzian action, including its prefactor.

In \cite{Maldacena:2016hyu}, the leading connected contribution to the four point function of fermions was computed by summing ladder diagrams. Taking a strict IR limit of the ladder diagrams produces a divergence.\footnote{In the dual gravity setup this phenomenon was elegantly discussed in~\cite{Almheiri:2014cka}.} The origin of this divergence is that reparametrisations of time is a spontaneously broken symmetry, and the associated Goldstone boson has zero action (due to the low dimensionality of the problem). Backing away from the IR leads to explicit breaking and makes time reparametrisations pseudo-Goldstone bosons. They contribute to the four point function through the propagator of small fluctuations  around the saddle  $\expval{\de f(u) \de f(0)}$, where $\de f(u)$ is defined through $f(u)=u+\de f(u)$. This propagator can be read off from the ladder diagrams, and it matches with the prediction of the linearisation of~\eqref{eqn:SchwarzianIntro}, with numerical input required to fix $\al_S$ for general $p$, and $\al_S={1\ov 4 p^2}$ for large $p$. In~\cite{Maldacena:2016hyu} a further symmetry argument is presented for the nonlinear completion to the full Schwarzian action~\eqref{eqn:SchwarzianIntro}.

In \cite{Kitaev:2017awl}, an alternative argument is presented for the nonlinear Schwarzian. The collective field formulation gives rise to a nonlocal action, which then is separated into two parts, one that is time reparatmetrisation invariant, and a deformation that breaks this symmetry. The latter term is large, but it is argued that it can be replaced by a small deformation in the IR regime (with $\al_S$ fixed by numerics), and the Schwarzian action is obtained this way. We regard our first argument as a close relative of the argument of \cite{Kitaev:2017awl}. However the large $p$ limit we consider is more controlled: the collective field formulation leads to a local conformal action (in two times), and it is only boundary conditions that break the reparametrisation symmetry. We can then use  boundary CFT (BCFT) renormalisation group technology to derive the Schwarzian action.

The outline of the paper is as follows. In Sec.~\ref{sec:SYKreview} we provide a brief review of the SYK model, highlighting the infrared regime and the emergence of a time reparametrisation soft mode and how its action breaks reparametrisation invariance. In Sec.~\ref{sec:largep} we then describe the large $p$ limit, which is characterised by a Liouville action with non-conformal boundary conditions. In particular, in Sec.~\ref{sec:BCFT} we show that the saddle point of this theory behaves like the one point function of a bulk vertex operator in a BCFT, deformed by an irrelevant boundary operator that we identify to be the displacement operator $\hat{D}$. Our first derivation consists of evaluating $\hat{D}$ on the reparametrisations of the saddle, finding that the deformation is, as expected, the Schwarzian. Our second derivation presented in Sec.~\ref{sec:Ansatz}, is less elegant, but more hands-on. We expect that this method can be generalised to other systems, where the Schwarzian is expected to emerge. It consists of building an Ansatz for the microscopic field configuration, and plugging it into the Liouville action. In particular, we look for a configuration that obeys the following criteria: its starting point is the reparametrisation of the thermal saddle restricted to obey the Kubo–Martin–Schwinger (KMS) condition and deformed to satisfy the non-conformal boundary conditions. Having found a configuration that follows the outlined criteria, we can compute its action: once again we recover the Schwarzian action. Finally, we study the SYK chain in Sec.~\ref{sec:chain}. We derive its soft mode action using both of our methods: from the perspective of conformal perturbation theory the inter-site coupling in the chain corresponds to the addition of a marginal operator to two decoupled Liouville BCFTs, while it is straightforward to make our Ansatz space-dependent and to evaluate the SYK chain action on it. 
Both methods lead to the same soft mode action that we call the Schwarzian chain following~\cite{Altland:2019lne}:
\es{eqn:chainIntro}{
    S=-\frac{N}{ 4 p^2}\sum_{x=0}^{M-1}&\Biggl[ (\pi\delta v)\int_0^{2\pi}d\tau\, \textrm{Sch}[\tan(f_x/2),\tau]+\\&+\alpha\int d\tau_1 d\tau_2\Biggl[  \frac{ \sqrt{f_x'\left(\tau _1\right)} \sqrt{f_x'\left(\tau _2\right)}}{\sqrt{\sin ^2\left(\frac{ f_x\left(\tau _1\right)-f_x\left(\tau
   _2\right)}{2}\right)}}   \frac{ \sqrt{f_{x+1}'\left(\tau _1\right)}\sqrt{ f_{x+1}'\left(\tau _2\right)}}{\sqrt{\sin ^2\left(\frac{ f_{x+1}\left(\tau _1\right)-f_{x+1}\left(\tau
   _2\right)}{2}\right)}} \Biggr] \Biggr]\,,
}
where $f_x$ is the reparametrisation degree of freedom for site $x$, $M$ is the number of lattice sites, and $\al$ is the inter-site coupling. Note that the inter-site coupling leads to an action is non-local in time, as was discussed before in~\cite{Maldacena:2018lmt,Altland:2019lne,Almheiri:2019jqq,Milekhin:2021sqd,Milekhin:2021cou}.

{\bf Note added:} During the final stages of our work we learned about an independent work by Berkooz, Frumkin, Mamroud, and Seitz~\cite{Berkooz:2024ifu} that also derives the nonlinear Schwarzian using a similar Ansatz to the one we use in our second derivation. We comment briefly on the differences between the two Ans\"atze in Sec.~\ref{sec:Ansatz}.

\section{Brief review of the SYK model}\label{sec:SYKreview}
The Sachdev-Ye-Kitaev model is an ensemble of quantum mechanical models. A member of the ensemble consists of N Majorana fermions with $p$-body interactions and is described by the following Hamiltonian:
\begin{equation}
    \label{eqn:SYK hamiltonian}
    H=\sum_{1\leq i_1\leq i_2\leq ... \leq i_p \leq N}J_{i_1i_2...i_p}\Psi_{i_1}...\Psi_{i_p},
\end{equation}
where the coupling $J_{i_1,...i_p}$ has a Gaussian distribution with
\begin{equation}
    \label{eqn: coupling average}
    \begin{split}
      \langle J_{i_1...i_p}\rangle&=0\,, \\  \langle J_{i_1...i_p}^2\rangle=\frac{J^2(p-1)!}{N^{p-1}}&=\frac{2^{p-1}}{p}\frac{\mathcal{J}^2(p-1)!}{N^{p-1}}\,.  
    \end{split}
\end{equation}
In the large $N$ limit with annealed disorder, we can realise the average over the disorder ensemble for $J_{i_1,...,i_p}$, by directly averaging the partition function \cite{Sachdev_1993}. We find the following action:
\es{eqn:SYK eff action}{
    I[G,\Sigma]=-\frac{1}{2}\textrm{log det}(\partial_{\tau}-\Sigma)+\frac{1}{2}\int\int(\Sigma G-\frac{1}{p}J^2G^p)\,.
}
The field $G$ is the Euclidean bilinear: 
\begin{equation}
    \label{eqn: G-definition}
    G(\tau)=\frac{1}{N}\sum_i T\Psi_i(\tau)\Psi_i(0)
\end{equation}
and $\Sigma$ is the self energy. Note that $G,\, \Sigma$ are fluctuating fields. 

The classical equations of motion derived by extremising (\ref{eqn:SYK eff action}) are the Schwinger-Dyson equations:
\begin{equation}
\label{eqn:Schwinger-Dyson}
\begin{split}
       G&=[\partial_{\tau}-\Sigma]^{-1}\,, \\
       \Sigma &=J^2G^{p-1}\,.
\end{split}
\end{equation}
The solutions to these equations give the leading large $N$ value for the propagator $\expval{G(\tau)}$  and self-energy.
It is easy to see that for zero coupling we get the propagator:
\begin{equation}
\label{eqn:free propagator}
\expval{G^\text{free}(\tau)}=\frac{1}{2}\textrm{sgn}(\tau)\,,
\end{equation}
whose small $\tau$ behaviour follows from the anticommutation relation of Majorana fermions, and hence  $G(\tau)\to \frac{1}{2}\textrm{sgn}(\tau)$ as $\tau\to 0$ will be imposed as a constraint in what follows.

\subsection{Low energy limit and the Schwarzian}
We are interested in studying the IR regime of the SYK model. We notice that, since $\Sigma$ is proportional to $J^2$, in the low energy limit the term $\partial_{\tau}$ in equation (\ref{eqn:Schwinger-Dyson}) is negligible and can be dropped. We can then write a new set of IR equations of motion:
\begin{equation}
\label{eqn: IR equations}
\begin{split}
    \int d\tau_2G(\tau,\tau_2)\Sigma(\tau_2,\tau_1)&=-\delta(\tau-\tau_1)\,,\\
    \Sigma(\tau,\tau_1)&=J^2G(\tau,\tau_1)^{p-1}\,.
\end{split}
\end{equation}
This set of equations is invariant under $\tau\rightarrow f(\tau)$, provided that the fields transform as
\begin{equation}
\label{eqn:conformal limit for G and Sigma}
\begin{split}
    G(\tau_1,\tau_2)&\rightarrow[f'(\tau_1)f'(\tau_2)]^{1/p}\,G(f(\tau_1)f(\tau_2))\,,\\
    \Sigma(\tau_1,\tau_2)&\rightarrow[f'(\tau_1)f'(\tau_2)]^{(p-1)/p}\,\Sigma(f(\tau_1)f(\tau_2))\,.
\end{split}
\end{equation}
This reparametrisation invariance is then explicitly broken once we take into account the $\partial_{\tau}$ term that we discarded in the IR limit. In particular, the violation of this symmetry was argued to be described by the following action~\cite{KitaevTalks,Maldacena:2016hyu,Kitaev:2017awl}: 
\begin{equation}
    \label{eqn:Schwarzian as in Lin}
    S=-\frac{N\alpha_S}{\mathcal{J}}\int du \, \textrm{Sch}[f,u]\,,
\end{equation}
where Sch$[f,u]$ is the Schwarzian derivative and $\alpha_s=\frac{1}{4p^2}$ in the large $p$ limit. In the following, we strengthen and generalise its derivation in the large-$p$ SYK model.  
%This result has been argued for by considering reparametrisations of the bilocal field $G$: we are interested in a derivation that starts instead from the microscopic description, which will then allow us to analyze the low energy limit in the SYK chain. 

\section{Liouville action}\label{sec:largep}
\begin{figure}
    \centering
    \begin{tikzpicture}
        \draw[fill,cambridgeblue!30] (0,3.5) -- (1.75,1.75) -- (3.5,3.5)--(1.75,5.25);
        \draw[thick,->] (0,0) -- (6,0) node[anchor=north west] {$\tau_1$};
        \draw[thick,ao(english),->] (1.75,1.75) -- (4.5,4.5) node[anchor=south west] {$\bar{\tau}$};
        \draw[thick,->] (0,0) -- (0,6) node[anchor=south east] {$\tau_2$};
        \draw[thick,ao(english),->] (1.75,1.75) -- (-1,4.5) node[anchor=south east] {$\delta \tau$};
        \draw[thick, azure(colorwheel)] (1.75,1.75) -- (3.5,3.5);
        \draw[thick, azure(colorwheel)] (0,3.5) -- (1.75,5.25);
        \draw[azure(colorwheel)] (3.5,-0.25) node[] {$2\pi$};
        \draw[azure(colorwheel)] (-0.35,3.5) node[] {$2\pi$};
        \draw[azure(colorwheel)] (1.75,-0.25) node[] {$\pi$};
        \draw[azure(colorwheel)] (-0.35,1.75) node[] {$\pi$};
        \draw[azure(colorwheel)] (-0.35,5.25) node[] {$3\pi$};
        \draw[dashed,azure(colorwheel)] (3.5,0) -- (3.5,3.5);
        \draw[dashed,azure(colorwheel)](0,3.5) -- (3.5,3.5);
        \draw[dashed,azure(colorwheel)] (1.75,0) -- (1.75,1.75);
        \draw[dashed,azure(colorwheel)](0,1.75) -- (1.75,1.75);
        \draw[dashed,azure(colorwheel)](0,5.25) -- (1.75,5.25);
        \draw[azure(colorwheel), fill] (0.5,3) circle(0.1cm);
        \draw[azure(colorwheel), fill] (3, 4) circle(0.1cm);
    \end{tikzpicture}
    \caption{Region of integration when $\beta$ is set to $2\pi$. Taking into account both KMS and the symmetry condition for $g$, we need to integrate only over the shaded region to get the correct action. Moreover, these conditions imply the identification of the two blue points $g(x,2\pi-x)=g(2\pi-x,2\pi+x)$.}
    \label{fig:region of integration} 
\end{figure}
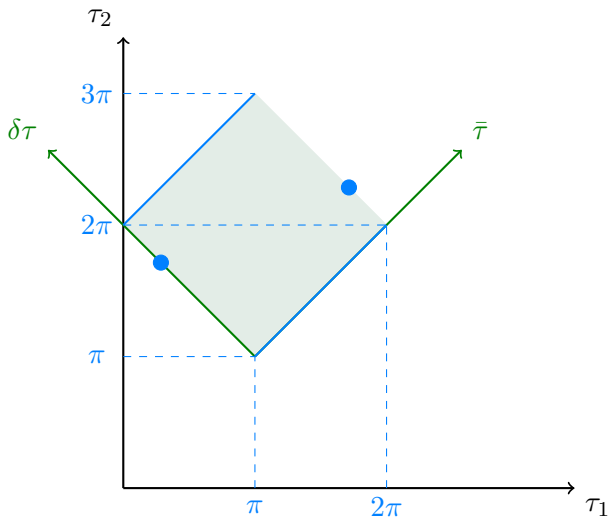

To achieve better analytical control, after the large $N$ limit we take the large $p$ limit. This order of limits means that the ratio $\lambda=\frac{2p^2}{N}$ is infinitesimal.\footnote{The double scaling limit in which $\lambda$ is held fixed is also very interesting~\cite{Erdos:2014zgc,Cotler:2016fpe,Berkooz:2018jqr}. However in the double scaling limit all modes of the collective field fluctuate strongly, hence the Schwarzian cannot dominate the dynamics.}  In order to do so, it is convenient to define the following field $g$:
\begin{equation}
    \label{eqn:small g-definition}
     G(\tau_1,\tau_2)=\frac{\textrm{sgn}(\tau_1-\tau_2)}{2}\left(1+\frac{g(\tau_1,\tau_2)}{p}\right).
\end{equation}
From this definition we see that, in order for (\ref{eqn:small g-definition}) to be consistent with the small time separation limit discussed below
 (\ref{eqn:free propagator}), we will need to enforce the boundary condition $g(\tau_1=\tau_2)=0$. Furthermore, since the bilinear should be antisymmetric for $\tau_1\rightarrow\tau_2$, $g$ needs to be symmetric: $g(\tau_1,\tau_2)=g(\tau_2,\tau_1)$. $g$ also needs to satisfy the KMS boundary conditions: $g(\tau_1,\tau_2+\beta)=g(\tau_1,\tau_2)$ and $g(\tau_1+\beta,\tau_2)=g(\tau_1,\tau_2)$. As shown in \cite{Cotler:2016fpe}, the action for $g$ is
\begin{equation}
    \label{eqn:liouville}
    I[g]=\frac{N}{4p^2}\int d\tau_1 d\tau_2\left[-\mathcal{J}^2e^{g(\tau_1,\tau_2)}+\frac{1}{4}\partial_{\tau_1}g(\tau_1,\tau_2)\partial_{\tau_2}g(\tau_1,\tau_2)\right],
\end{equation}
which has the form of a Liouville action. Enforcing both KMS and the symmetry condition for $g$ restricts our region of integration to the shaded diamond in figure (\ref{fig:region of integration}), with the boundary condition $g(\tau_1=\tau_2)=0=g(\tau_1=\tau_2-\beta)$ and the identification $g(x,\beta-x)=g(\beta-x,\beta+x)$, and with the prefactor in equation (\ref{eqn:liouville}) now becoming $\frac{N}{2p^2}$. Since this prefactor is large, it is useful to analyse the saddle point:
\begin{equation}
    \label{eqn:saddle for Liouville}
    e^{g_{*}(\tau_1,\tau_2)}=\left(\frac{\cos\left[\frac{\pi v}{2}\right]}{\cos\left[\pi v\left(\frac{1}{2}-\frac{|\tau_1-\tau_2|}{\beta}\right)\right]}\right)^2 \,,
\end{equation}
where $v$ is defined through the equation:
\es{vdef}{
\beta\mathcal{J}=\frac{\pi v}{\textrm{cos}\frac{\pi v}{2}}\,.
}
Note that the low energy limit, $\beta\mathcal{J}\gg1$, corresponds to the regime where $v\rightarrow1$. It will then be convenient to redefine $v=1-\delta v$ and then take $\delta v\rightarrow0$. We will also work with the following set of coordinates:
\es{eqn: delta tau and tau bar}{
    \delta\tau&=\tau_2-\tau_1\,,\\
    \bar{\tau}&=\frac{\tau_2+\tau_1}{2}\,.\\
}
It is convenient for us to work with a rescaled field, $\gamma$:
\begin{equation}
    \label{eqn:gamma-definition}
e^{\gamma(\delta\tau,\bar{\tau})}=\le({\beta\mathcal{J}\ov 2\pi}\ri)^2e^{g(\delta\tau,\bar{\tau})}.
\end{equation}
From now on we set $\beta=2\pi$, which can be reinstated using dimensional analysis.
The action for $\gamma$ can be derived from (\ref{eqn:liouville}):
\begin{equation}
    \label{eqn:liouville for gamma}
    I[\gamma]=\frac{N}{2p^2}\int d\tau_1 d\tau_2\left[-e^{\gamma(\tau_1,\tau_2)}+\frac{1}{4}\partial_{\tau_1}\gamma(\tau_1,\tau_2)\partial_{\tau_2}\gamma(\tau_1,\tau_2)\right].
\end{equation}
Note that in equation (\ref{eqn:liouville for gamma}), the explicit dependence on $\mathcal{J}$ has been reabsorbed into $\gamma$: now it only appears in the boundary conditions
\begin{equation}
    \label{eqn:bdy conditions for gamma}
    \gamma(0,\bar{\tau})=\textrm{log}[\mathcal{J}^2].
\end{equation}
%\MB{would it be more effective if I wrote down the bdy conditions for $e^\gamma$ instead?}
The saddle point for this field is
\begin{equation}
   \label{eqn:saddle point for gamma}
   e^{\gamma_*(\delta\tau,\bar{\tau})}=\left(\frac{v}{2\cos\left[ v\left(\frac{\pi-\delta\tau}{2}\right)\right]}\right)^2.
\end{equation}
\subsection{Reparametrisations for the Liouville action}
\label{section:reparametrisation modes for the liouville action}
We note that the Liouville action is invariant under reparametrisations of the form 
\begin{equation}
    \tau_1\rightarrow f(\tau_1)\,,\qquad
    \tau_2\rightarrow h(\tau_2)\,,
\end{equation}  
provided that:
\begin{equation}
    \label{eqn: reparametrisation invariance for small g}
    \gamma(\tau_1,\tau_2)\rightarrow \gamma(\tau_1,\tau_2)+\textrm{log}\left[f'(\tau_1)\right]+\textrm{log}\left[h'(\tau_2)\right].
\end{equation}
This statement is equivalent to (\ref{eqn:conformal limit for G and Sigma}).

Let us start by looking at the reparametrised saddle:
\begin{equation}
    \label{eqn:reparametrized saddle}
    e^{\tilde \gamma}=\frac{v^2}{4\cos^2\left[v\frac{\pi+f(\tau_1)-h(\tau_2)}{2}\right]}f'(\tau_1)h'(\tau_2).
\end{equation}
Imposing $\gamma(\tau_1,2\pi-\tau_1)=\gamma(2\pi-\tau_1,2\pi+\tau_1)$ results in the following set of equations:
\es{eqn: f and g}{
        -f(\tau_1)+h(2\pi-\tau_1)&=f(2\pi-\tau_1)-h(2\pi+\tau_1)+2\pi\,,\\
        \quad f'(\tau_1)h'(2\pi-\tau_1)&=f'(2\pi-\tau_1)h'(2\pi+\tau_1),
}
which in turn imply 
\es{eqn: winding for f}{
        f(\tau_1)&=h(\tau_1)\,,\\
        f(\tau_1+2\pi)&=f(\tau_1)+2\pi\,,\\
        f'(\tau_1+2\pi)&=f'(\tau_1)\,.
}
Thus we are left with one reparametrisation symmetry. This is also explicitly broken by the boundary condition $\gamma(\tau_1=\tau_2)=\textrm{log}[\mathcal{J}^2].$ However, in the strong coupling $\mathcal{J}\to\infty$ limit the boundary condition is conformal and the symmetry is restored. This fact motivates our first approach.

\section{Boundary CFT approach to the Schwarzian}\label{sec:BCFT}
We can give a short, abstract derivation of the Schwarzian action based on boundary CFT (BCFT). 
The Liouville theory in \eqref{eqn:small g-definition} is a CFT. Here we are considering it on a Mobius strip, equipped with non-conformal boundary conditions
\es{nonconfBC}{
e^{\gamma(\bar \tau,\de \tau=0)}=\mathcal{J}^2\,.
}
Below we explain that for small $\delta v$ this boundary condition is close to the ZZ conformal boundary condition,\footnote{The conformal boundary conditions four Liouville theory have been classified, besides ZZ there is a one parameter family of FZZT boundary conditions.} and we can account for the difference using (boundary) conformal perturbation theory.

The ZZ BCFT on the half space $\de\tau>0$ (and $\bar{\tau}$ unrestricted) is defined by the boundary condition 
\es{confBC}{
e^{\gamma}\sim \frac{1}{\de \tau^2}\, .
}
The behaviour of the saddle point \eqref{eqn:saddle point for gamma} in the regime $\delta v\lesssim \de \tau\ll \beta$ is:
\es{saddleExp}{
e^{\gamma_*(\bar \tau,\de \tau)}&=\<e^\gamma\>= \frac{1}{ (\pi \delta v+\de \tau)^2}+\dots\\
&=\frac{1}{ \de \tau^2}-\frac{2\pi \delta v}{ \de \tau^3}+\dots\,.
}
 We observe that this is the one point function of the bulk vertex operator $e^\gamma$ in the ZZ BCFT deformed by an irrelevant operator, since the perturbation grows towards the boundary $\de \tau \to 0$ and is negligible for large $\de \tau$. 

Let us denote the irrelevant boundary operator as $\hat {\cal O}$. We can then write
\es{ConfPert}{
\<e^\gamma\>&=\<e^\gamma\>_\text{ZZ}+\lam \int d\bar\tau \, \<\hat {\cal O}(\bar\tau) e^\gamma\>_\text{ZZ}+\dots\\
&=\<e^\gamma\>_\text{ZZ}+\lam \int d\bar\tau \, \frac{\alpha}{ \de \tau^{2-\hat\De}\, (\de \tau^2+\bar \tau^2)^{\hat\De}}+\dots\\
&= \frac{1}{ \de \tau^2}+ \frac{\lam \alpha' }{ \de \tau^{1+\hat\De} }+\dots
\,,
}
where in the second line we used the general form of a boundary bulk two point function from \cite{McAvity:1995zd} and in the third line we have absorbed some $\hat\De$ dependent factors into $\alpha'$. We conclude that we are looking for a boundary operator with dimension $\hat\De=2$. 

Either from this computation or from simply looking at the first line of \eqref{saddleExp} we identify $\hat{\cal O}=\hat D$, the displacement operator. The displacement operator has a nice geometric action: it locally moves the boundary inwards by a unit distance. Since the ZZ boundary was moved outwards by $\pi \delta v$, we find that $\lam=-\pi \delta v$. We conclude that we are studying the deformed BCFT\footnote{It would be interesting to verify this result in the double scaled SYK model, where the Liouville theory is in the quantum regime by matching its free energy with that of the SYK model computed in~\cite{Cotler:2016fpe,Berkooz:2018jqr}.}
\es{deformedAction}{
S_\text{ZZ}-\pi \delta v \int d\bar\tau \,\hat D(\bar\tau)+\dots\,.
}

Liouville theory with ZZ boundary conditions has an infinite set of saddle points, since the reparametrisations of $\gamma_*$ (see section \ref{section:reparametrisation modes for the liouville action}),
\begin{equation}
    \label{eqn:reparametrized saddle2}
    e^{\gamma^{(f)}}=\frac{ f'(\tau_1)f'(\tau_2)}{\sin^2\le(\frac{f(\tau_1)-f(\tau_2)}{ 2}\ri)}
\end{equation}
also satisfy the ZZ boundary conditions. Hence they have the same action as $\gamma_*$. To first order in perturbation theory, we then only need to evaluate $\hat D(\bar\tau)$ on this family of saddle points. To do so, we recall that $\hat D=T^{\de \tau \de \tau}\vert_{\de \tau=0}$, where $T^{ab}$ is the stress tensor.\footnote{When working on the diamond shaped fundamental region, there is of course another boundary at $\de \tau=2\pi$ whose contribution we have to take into account.} Using the classic result from~\cite{Dorn:2006ys} that the Liouville stress tensor evaluated on a saddle $\gamma$ is
\es{Texpr}{
T_{11}(\tau_1)=\frac{N}{ 4p^2}\,e^{\gamma/2}\p_{11} e^{-\gamma/2}\,, \qquad T_{22}(\tau_2)=\frac{N}{ 4p^2}\,e^{\gamma/2}\p_{22} e^{-\gamma/2}\,,
}
and plugging in the saddle \eqref{eqn:reparametrized saddle2}, we obtain that\footnote{This calculation was done already in~\cite{Dorn:2006ys}.} 
\es{Texpr2}{
T_{11}(\tau)=T_{22}(\tau)=\frac{N}{ 8p^2}\,\textrm{Sch}[\tan(f/2),\tau]\,.
}
Transforming these components into $T^{\de \tau \de \tau}=T_{11}+T_{22}$\footnote{This equation holds, since the tracelessness of the stress tensor in the light cone coordinates $\tau_{1,2}$ is written as $T_{12}=0$.} and evaluating at ${\de \tau=0}$, we get that
\begin{equation}
\label{eqn:deformedAction2}
   -\pi \delta v \int d\bar\tau \,\hat D(\bar\tau)\big\vert_{\gamma^{(f)}}=-\frac{N \pi \delta v}{ 4p^2}\int d\bar\tau \,\textrm{Sch}[\tan(f/2),\bar{\tau}]\,, 
\end{equation}
which is, as expected just the thermal Schwarzian, $\textrm{Sch}[\tan(f/2),\tau]$. We now need to compare $\frac{N \delta v\pi}{4p^2}$ to $\frac{N \alpha_s}{\mathcal{J}}=\frac{N }{4p^2\mathcal{J}}$ from equation (\ref{eqn:Schwarzian as in Lin}). From eq (\ref{eqn:saddle for Liouville}), we can see that, in the IR limit, $\frac{1}{\mathcal{J}}=\pi \delta v+O(\delta v)^2$: it is then clear that the two coefficients are the same. Note that from the $\de\tau=0$ boundary we get an integral over $\bar\tau\in(\pi,2\pi)$, while from the $\de\tau=2\pi$ boundary we get an integral over $\bar\tau\in(0,\pi)$: together they complete the thermal circle.

We conclude that the infinite set of saddle points \eqref{eqn:reparametrized saddle2} are lifted at linear order in $\de v$. The set of field configurations is vastly larger than this soft direction. However, the orthogonal ``hard directions" have an action that is $O(N/p^2)$, which is large in the large-$p$ SYK model. The fluctuations in these directions can then be set to their saddle point value zero.\footnote{These fluctuations then contribute an $O(1)$ amount to the free energy through their functional determinant.} Fluctuations in the reparametrisation direction in field space are enhanced by their small action (relative to other modes), and become $O(1)$ in the ultra-low temperature regime $\de v=O(p^2/N)$, where their action is $O(1)$.

\section{An alternative derivation for the Schwarzian}\label{sec:Ansatz}
As outlined in section \ref{section:reparametrisation modes for the liouville action}, the Liouville action is invariant under independent reparametrisations of $\tau_1$ and $\tau_2$, which then get restricted to one reparametrisation by the KMS condition. The remaining diagonal reparametrisation symmetry is broken by the boundary conditions at $\tau_1=\tau_2$. 
%\begin{equation}
%\begin{cases}
%    \tau_1\rightarrow f(\tau_1)\\
%    \tau_2\rightarrow h(\tau_2)\\
%    \end{cases}. 
%\end{equation}
The goal for this section is to find the action that describes the behaviour of the reparameterisation modes, starting from~\eqref{eqn:liouville for gamma}: this is an alternative way of deriving the Schwarzian action. 
%that can also be generalised to the SYK chain. 
In order to perform this computation, we first want to find a field configuration that includes reparametrisations of the saddle point and that obeys both the boundary conditions and KMS.

\subsection{Field configuration}

\begin{figure}[ht]
    \centering
    \begin{subfigure}{.48\textwidth}
       \centering
        \includegraphics[width=.9\linewidth]{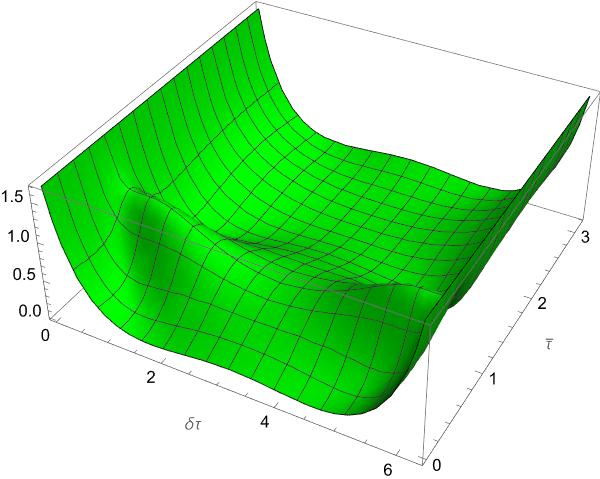}
        \caption{Field Configuration for $v=0.8$}
    \end{subfigure}
    \begin{subfigure}{.48\textwidth}
       \centering
        \includegraphics[width=.9\linewidth]{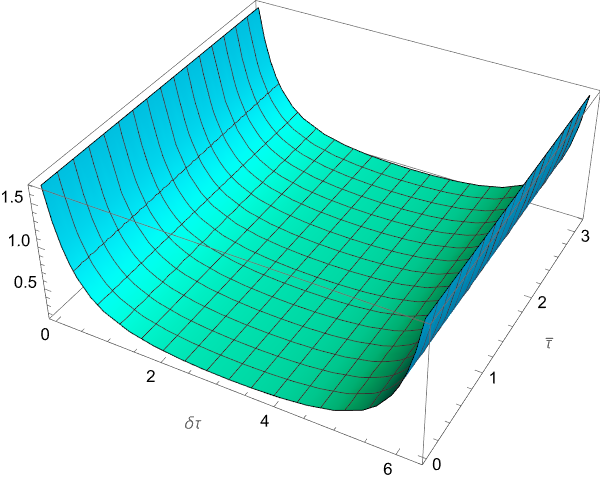}
    \caption{Saddle Point solution for $v=0.8$}
    \end{subfigure}
%    \begin{subfigure}{.5\textwidth}
%        \includegraphics[width=.8\linewidth]{plots/rep9.png}
%        \caption{Field Configuration for $v=0.9$}
%    \end{subfigure}
%    \begin{subfigure}{.5\textwidth}
%        \includegraphics[width=.8\linewidth]{plots/saddle9.png}
%        \caption{Saddle Point solution for $v=0.9$}
%    \end{subfigure}
%    \begin{subfigure}{.5\textwidth}
%        \includegraphics[width=.8\linewidth]{plots/rep99.png}
%        \caption{Field Configuration for $v=0.99$}
%    \end{subfigure}
%    \begin{subfigure}{.5\textwidth}
%        \includegraphics[width=.8\linewidth]{plots/saddle99.png}
%        \caption{Saddle Point solution for $v=0.99$}
%    \end{subfigure}
    \caption{Three dimensional plots of $e^{\ga_f(\tau_1,\tau_2)}$ and $e^{\ga_*(\tau_1,\tau_2)}$ for $v=0.8$. On the left we have our Ansatz for the field configuration when $f(\tau)=\tau +0.1 \sin (2 \tau )+0.2 \cos (3 \tau )$ and on the right we are plotting the saddle point solution. By construction, both $e^{\ga_f(\tau_1,\tau_2)}$ and $e^{\ga_*(\tau_1,\tau_2)}$ are equal to $\mathcal{J}^2$ on the boundary. %Furthermore we note that $\left.e^{\ga(\tau_1,\tau_2)}\right|_f$ does not diverge for $v\rightarrow1$.
    }
    \label{fig:field configurations}
\end{figure}

We start from the reparametrised saddle~\eqref{eqn:reparametrized saddle}, but with the constraint $ f(\tau_1)=h(\tau_1)$ imposed by the KMS symmetry discussed in \eqref{eqn: winding for f}:
\begin{equation}
    \label{eqn:reparametrized saddle3}
   e^{\tilde \ga}=\frac{v^2}{4\cos^2\left[v\frac{\pi+f(\tau_1)-f(\tau_2)}{2}\right]}f'(\tau_1)f'(\tau_2)\,.
\end{equation}
This field configuration violates the boundary conditions. However, if we take the $v\to 1$ limit, we get that the violation is $f$ independent:
\begin{equation}
    \label{eqn:reparametrized saddle4}
   e^{\tilde \ga}\big\vert_{v\to 1}=\frac{f'(\tau_1)f'(\tau_2)}{4\sin^2\left[\frac{f(\tau_1)-f(\tau_2)}{2}\right]}\approx {1\over \de \tau^2}\,.
\end{equation}
(This is the same equation as \eqref{eqn:reparametrized saddle2}.) If we divide this with the $v=1$ saddle, we get a field configuration that goes to $1$ at the boundary:
\begin{equation}
    \label{eqn:reparametrized saddle5}
  \frac{\sin ^2\left(\frac{\tau _2-\tau _1}{2} \right)}{\sin^2\left[\frac{f(\tau_1)-f(\tau_2)}{2}\right]}\,f'(\tau_1)f'(\tau_2)\,.
\end{equation}
We then multiply this with the $v\neq 1$ saddle to enforce the boundary conditions and produce our Ansatz  (see figure~\ref{fig:field configurations}):
\begin{equation}
    \label{eqn:final field configuration for reparam}
   e^{\gamma_f(\tau_1,\tau_2)}\equiv\frac{v^2}{4\cos^2\left[ v\left(\frac{\pi+\tau_1-\tau_2}{2}\right)\right]}\frac{\sin ^2\left(\frac{\tau _2-\tau _1}{2} \right)}{ \sin^2\left(\frac{f\left(\tau _2\right)-f\left(\tau _1\right)}{2}
   \right)} f'\left(\tau _1\right) f'\left(\tau _2\right)\,.
\end{equation}
This Ansatz is designed to embed the soft reparametrisation mode into the microscopic field configuration $\ga(\tau_1,\tau_2)$, and hence capture the soft mode. This mode is special, a generic configuration instead has a large $O(N/p^2)$ action, and fluctuations in the hard directions can be set to zero.

Admittedly, our construction of the Ansatz is ad hoc. However, it satisfies all conditions on the field configuration, and it is a ``small deformation" of the $v=1$ family of saddle points. Also, for linearised reparametrisations $f(u)=u+\ep e^{-i n \tau}$ it gives:
\es{LinRepar}{
e^{-\gamma_*(\delta\tau,\bar{\tau})} e^{\gamma_f(\delta\tau,\bar{\tau})}&=1+2i\ep\, f_n(\de \tau) e^{-i n \bar\tau}+O(\lam^2)\,,\\
f_n(\de \tau)&={\sin{n \de \tau\ov 2}\ov \tan{ \de \tau\ov 2}}-n \cos{n \de \tau\ov 2}\,.
}
The $f_n(\de \tau)$ functions are identical to those defined in eq.~(3.109) of~\cite{Maldacena:2016hyu}. They realise infinitesimal reparametrisations of the $v=1$ saddle. We conclude that our Ansatz is accurate to linear order. At nonlinear order, the goal is to capture the soft direction in field space. The potential issue with the Ansatz could be that at nonlinear order in $\de f$ it mixes with the ``hard" directions. The output of our calculation will be an action that is small, $O(\de v)$, which a posteriori confirms that there is no large mixing with the hard directions. However, an $O(\de v)$ action is also consistent with an $O(\sqrt{\de v})$ mixing with hard directions. We can verify that this does not happen in the region $\de \tau\gg \de v$, where our Ansatz behaves as
\begin{equation}
    \label{eqn:expansion of e^g around delta t=0}
     \begin{split}
     e^{\gamma_f(\delta\tau,\bar{\tau})}=&e^{\gamma_*}\le[1+{\de \tau^2 \ov 6}\textrm{Sch}\le[\tan\le(f/ 2\ri),u\ri]+O\left(\delta \tau ^3\right) \ri]\,, 
     \end{split}
\end{equation}
which is the expected IR behaviour of the soft mode. We conclude that in the worst case the Ansatz deviates from the true soft mode in the region $\de \tau\sim \de v$, which can only result in $O(\de v)$ mixing, which is negligible. Indeed, we obtain the Schwarzian action from this Ansatz, which we know, from our first method and earlier literature, is the correct result.

The mixing issue could be further investigated by linearising around the Ansatz with a finite $f$ and finding that the zero mode of the linearised equation is indeed in the direction of the Ansatz with $f+\de f$; we leave this computation for the future. 

{\bf Note added:}  We briefly comment on the parallel work~\cite{Berkooz:2024ifu}. The authors consider an Ansatz distinct, but close to ours in spirit. Their Ansatz for linearised $f(u)$ gives ${\cal F}_n(\de \tau)$, which are the generalisation of $f_n(\de \tau)$ for finite $\de v$ and reduces to them as $\de v\to 0$.\footnote{Consider the finite $\de v$ saddle and linearise around it. The eigenfunctions of the resulting differential operator are $\psi_{m}(\de \tau) e^{-i n\bar\tau}$ with eigenvalue proportional to $n^2-m^2$, where $n\in \mathbb{Z}$ and $m$ is a solution of a transcendental equation~\cite{Choi:2019bmd}. Let us denote the $m$ closest to $n$ as $m(n)$: this gives the softest eigenmode. For small  $\de v$ we get $m(n)=n(1-\de v+\dots)$. We then define ${\cal F}_n(\de \tau)=\psi_{m(n)}(\de \tau)$.} This difference is immaterial for getting the correct soft mode action at $O(\de v)$, since it amounts to a negligible $O(\de v)$ mixing with hard modes. The important difference between our Ansatz and that of  \cite{Berkooz:2024ifu}, is that the latter only satisfies the boundary conditions approximately at small $\de v$, which the authors remedy by adding a boundary term to their action.

%Rewriting equation (\ref{eqn:final field configuration for reparam}) using the variables (\ref{eqn: delta tau and tau bar}) and expanding around $\delta\tau=0$, we find
%\begin{equation}
%     \label{eqn:expansion of e^g around delta t=0}
%      \begin{split}
%      \left.e^{\gamma(\delta\tau,\bar{\tau})}\right|_f=&\frac{v^2}{4\cos ^2\left(\frac{\pi  v}{2}\right)}  -   \frac{2 v^3 \sin ^4\left(\frac{\pi  v}{2}\right)}{ \sin ^3(\pi 
%    v)}\delta \tau+\frac{v^2\left(3 v^2+9
%    v^2 \tan ^2\left(\frac{\pi  v}{2}\right)-1\right) }{48 \cos ^2\left(\frac{\pi  v}{2}\right)
%   }\delta \tau ^2+\\&+\frac{ v^2  \left(-3 f''(\bar{\tau })^2+f'(\bar{\tau })^4+2 f^{(3)}(\bar{\tau}) f'(\bar{\tau })\right)}{48 \cos ^2\left(\frac{\pi  v}{2}\right)
%    f'(\bar{\tau})^2}\delta \tau ^2+O\left(\delta \tau ^3\right).   
%      \end{split}
% \end{equation}
% Most of this expansion is reparametrisation independent, with the only exception of the last term, which is proportional to the Schwarzian: this is consistent with the expansion for $G$ computed in \cite{Kitaev:2017awl}.

 Before we proceed it is important to emphasise the following issue. We are interested in considering the limit for $v\rightarrow1$: let us start by rewriting the saddle point solution as:
\begin{equation}
    \label{eqn:saddle pt with dv}
    e^{\gamma_*(\delta\tau,\bar{\tau})}=\frac{1}{4\textrm{cos}^2\left[ (1-\delta v)\left(\frac{\pi-\delta\tau}{2}\right)\right]}\,.
\end{equation}
For finite values of $\delta\tau$, taking the limit for $\delta v$ infinitesimal is straightforward:
\begin{equation}
    \label{eqn:small dv}
    e^{\gamma_*(\delta\tau,\bar{\tau})}\sim\frac{1}{4\textrm{cos}^2\left[\frac{\pi-\delta\tau}{2}\right]}\,.
\end{equation}
However, when we are close to the boundary, i.e.~when $\delta\tau\sim \delta v$, things become more delicate. As in~\eqref{saddleExp} we find
\begin{equation}
    \label{eqn:small dt small dv}
    e^{\gamma_*(\delta\tau,\bar{\tau})}\sim\frac{1}{(\delta v\pi+\delta\tau)^2}\,.
\end{equation}
As a result, in the following computation we will consider contributions from the bulk and from the boundary separately, paying close attention to the regime in which $\delta\tau\sim \delta v$.\\
 
\subsection{Near-boundary contributions}
To find the action for the reparametrisations, we plug (\ref{eqn:final field configuration for reparam}) into the Liouville Lagrangian: the expression we find is not particularly illuminating, so we will not report it.
As discussed, we need to consider contributions from the near-boundary region separately: we thus perform two rescalings, $\delta\tau\rightarrow\delta\tau' \delta v$ and $\delta\tau\rightarrow 2\pi-\delta\tau' \delta v$, and we expand around $\delta v\rightarrow0$. We find:
\begin{equation}
    \label{eqn:border contributions}
    \begin{split}
        S_{\textrm{bdy}}(\Lambda)=&\frac{N}{2p^2}\int_0^{\pi}d\bar{\tau}\int_0^\Lambda d\delta\tau'\frac{\delta \tau ' (\delta \tau '+2 \pi ) \left(-3 f''(\bar{\tau})^2+f'(\bar{\tau})^4+2
   f^{(3)}(\bar{\tau}) f'(\bar{\tau})\right)}{12 (\delta \tau '+\pi )^2 f'(\bar{\tau})^2}\delta v  \\
   +&\frac{N}{2p^2}\int_0^{\pi}d\bar{\tau}\int_0^\Lambda  d\delta\tau'  \frac{\delta \tau ' (\delta \tau '+2 \pi ) \left(-3 f''(\bar{\tau}+\pi )^2+f'(\bar{\tau}+\pi )^4+2 f^{(3)}(\bar{\tau}+\pi ) f'(\bar{\tau}+\pi )\right)}{12 (\delta \tau '+\pi )^2 f'(\bar{\tau}+\pi )^2}\delta v\,,
    \end{split}
\end{equation}
where $\Lam$ is a large cutoff in the rescaled time $\de \tau'$.
Integrating in $\delta\tau'$ we get:
\begin{equation}
    \label{eqn:integrated border contributions} 
    \begin{split}
        S_{\textrm{bdy}}(\Lambda)=&\frac{N}{2p^2}\int_0^{\pi}d\bar{\tau}\frac{\Lambda ^2 \left(-3 f''(\bar{\tau})^2+f'(\bar{\tau})^4+2 f^{(3)}(\bar{\tau}) f'(\bar{\tau})\right)}{12 (\Lambda +\pi ) f'(\bar{\tau})^2} \delta v\\
   +&\frac{N}{2p^2}\int_0^{\pi}d\bar{\tau}\frac{\Lambda ^2 \left(-3 f''(\bar{\tau}+\pi )^2+f'(\bar{\tau}+\pi )^4+2 f^{(3)}(\bar{\tau}+\pi )
   f'(\bar{\tau}+\pi )\right)}{12 (\Lambda +\pi ) f'(\bar{\tau}+\pi )^2} \delta v\,.
    \end{split}
\end{equation}
Looking at the second contribution, by performing a change of variable $\bar{\tau}\rightarrow\bar{\tau}-2\pi$ we can rewrite it as:
\begin{equation}
    \label{eqn:boundary-second contribution}
    \frac{N}{2p^2}\int_{\pi}^{2\pi}d\bar{\tau}\frac{\Lambda ^2 \left(-3 f''(\bar{\tau})^2+f'(\bar{\tau})^4+2 f^{(3)}(\bar{\tau}) f'(\bar{\tau})\right)}{12 (\Lambda +\pi ) f'(\bar{\tau})^2} \delta v\,.
\end{equation}
Adding the two contributions together we get
\begin{equation}
    \label{eqn:final boundary contribution}
    S_{bdy}(\Lambda)=\frac{N}{2p^2}\int_0^{2\pi}d\bar{\tau}\frac{\Lambda ^2 \left(-3 f''(\bar{\tau})^2+f'(\bar{\tau})^4+2 f^{(3)}(\bar{\tau}) f'(\bar{\tau})\right)}{12 (\Lambda +\pi ) f'(\bar{\tau})^2} \delta v\,.
\end{equation}
Note that the action in equation (\ref{eqn:final boundary contribution}) depends on the presence of a cutoff, $\Lambda$, and is divergent for $\Lambda\rightarrow\infty$; this divergence, however, will be cured once we consider contributions from the bulk.
\subsection{Bulk contributions}
\label{section:bulk contribution}
To compute the bulk contributions we can directly expand our Lagrangian in $\delta v$ as ${\cal L}_{bulk}={\cal L}_{bulk,0}+\de v {\cal L}_{bulk,1}+\dots$ and consider the contributions order by order. When integrating, to avoid double counting we will include a cutoff by considering only the interval $\delta\tau\in[\Lambda \delta v, 2\pi-\Lambda \delta v]$. Since the cutoff depends on $\de v$, when integrating the term ${\cal L}_{bulk,n}$, we expect to get contributions at every order $\de v^m$ with $m\geq n$.  First, we integrate ${\cal L}_{bulk,0}$, and after some lengthy manipulations (see appendix \ref{appendix:bulk contributions}) we find:
\begin{equation}
    \label{eqn:order zero relevant contrib}
    S_{bulk,0}(\Lambda)=\frac{N}{2p^2}\int_0^{2\pi}d\tau\left[-\frac{\Lambda  \left(-3 f''(\tau )^2+f'(\tau )^4+2 f^{(3)}(\tau ) f'(\tau )\right)}{12 f'(\tau )^2}\delta v\right]\,.
\end{equation}
Adding this contribution to eq (\ref{eqn:final boundary contribution}), we get:
\begin{equation}
    \label{eqn:order zero +boundary lambda}
    S_{bulk,0}(\Lambda)+S_{bdy}(\Lambda)=\frac{N}{2p^2}\int_0^{2\pi}d\tau\left[-\frac{\Lambda \pi  \left(-3 f''(\tau )^2+f'(\tau )^4+2 f^{(3)}(\tau ) f'(\tau )\right)}{12(\Lambda+\pi) f'(\tau )^2}\delta v\right].
\end{equation}
Taking the limit for $\Lambda\rightarrow\infty$ yields
\begin{equation}
    \label{eqn:order zero +boundary}
    S_{bulk,0}+S_{bdy}=\frac{N}{2p^2}\int_0^{2\pi}d\tau\left[-\frac{ \pi  \left(-3 f''(\tau )^2+f'(\tau )^4+2 f^{(3)}(\tau ) f'(\tau )\right)}{12 f'(\tau )^2}\delta v\right].
\end{equation}
As expected, adding contributions from the bulk cures the divergence that we had from the boundary. It is reasonable to expect that the cutoff dependence cancels at higher orders in $\delta v$ as well, and we have explicitly verified this to second order in  $\delta v$.
%for example, at second order in $\delta v$ we will have a contribution from $S_{bulk,0}(\Lambda)$ (by expanding up to second order in the cutoff) and one from $S_{bulk,1}(\Lambda)$ (by considering the first order contribution from the cutoff expansion). These two contributions, when added to the boundary action, will cure the $\Lambda$ divergence at subleading order. 

Next, when integrating ${\cal L}_{bulk,1}$ we can set the cutoff to zero, since keeping it would only result in subleading contributions. After some manipulations (see again appendix \ref{appendix:bulk contributions}), we can rewrite the contribution at first order as:
\begin{equation}
\label{eqn:first order nice}
    \de v\, S_{bulk,1}=\frac{N}{2p^2}\int_0^{2\pi}d\tau\left(-\frac{\pi   \left(-3 f''(\tau )^2+f'(\tau )^4+2 f^{(3)}(\tau ) f'(\tau )\right)}{6 f'(\tau )^2}\delta v\right).
\end{equation}
Adding everything together we get:
\begin{equation}
\label{eqn:adding contributions}
    S=-\frac{N}{2p^2}\frac{\delta v\pi}{2}\int_0^{2\pi}d\tau\left(\frac{1}{2} f'(\tau )^2+\frac{f^{(3)}(\tau )}{f'(\tau )}-\frac{3 f''(\tau )^2}{2 f'(\tau )^2}\right)=-\frac{N \delta v\pi}{4p^2}\int_0^{2\pi}d\tau\, \textrm{Sch}[\tan(f/2),\tau]\,,
\end{equation}
which is in complete agreement with (\ref{eqn:deformedAction2}).

\section{The nearest neighbour coupling in the SYK chain at low temperatures} \label{sec:chain}

The SYK chain consists of a chain of coupled SYK sites, with nearest neighbour interactions~\cite{Gu:2016oyy}. The Hamiltonian has the form:
\begin{align}
    \label{eqn:SYK chain hamiltonian}
    H=\sum_{x=1}^M\biggl(&\sum_{1\leq i_1\leq i_2\leq ... \leq i_p \leq N}J_{x,i_1i_2...i_p}\Psi_{i_1,x}...\Psi_{i_p,x}+\nonumber\\&\sum_{\substack{1\leq i_1\leq i_2\leq ... \leq i_{p/2} \leq N\\ 1\leq j_1\leq j_2\leq ... \leq j_{p/2} \leq N}}J'_{x,i_1...i_{p/2}j_1...j_{p/2}}\Psi_{i_1,x}...\Psi_{i_{p/2},x}\Psi_{j_1,x+1}...\Psi_{j_{p/2},x+1}\biggr)\,,
\end{align}
with:
\begin{equation}
    \label{eqn:constants for the chain}
    \langle J^2_{x,i_1,..i_p}\rangle=\frac{(p-1)!}{N^{p-1}}J_0^2,\quad \langle J'^2_{x,i_1...i_{p/2}j_1...j_{p/2}}\rangle=\frac{\left[(p/2)!\right]^2}{pN^{p-1}}J_1^2, \quad \mathcal{J}^2_{0,1}\equiv\frac{p}{2^{p-1}}J^2_{0,1}\,.
\end{equation}
We will identify $\mathcal{J}=\mathcal{J}_0$ from previous sections and define $\alpha\equiv\frac{\mathcal{J}_1^2}{\mathcal{J}^2}$.\footnote{This notation deviates from the convention of the literature, where the definition $\mathcal{J}^2\equiv\mathcal{J}_0^2+\mathcal{J}_1^2$ is used. Since we will take $\mathcal{J}_1\ll\mathcal{J}_0$, this is an unimportant difference.} In the large $p$ limit we can write the action as~\cite{Gu:2016oyy,Choi:2020tdj}:
\begin{equation}
    \label{eqn:liouville chain}
    I[g]=\frac{N}{4p^2}\sum_{x=0}^{M-1}\int d\tau_1 d\tau_2\left[-\mathcal{J}^2e^{g_x(\tau_1,\tau_2)}-\mathcal{J}_1^2e^{\frac{1}{2}(g_x(\tau_1,\tau_2)+g_{x+1}(\tau_1,\tau_2))}+\frac{1}{4}\partial_{\tau_1}g_x(\tau_1,\tau_2)\partial_{\tau_2}g_x(\tau_1,\tau_2)\right]\,.
\end{equation}

From the perspective of conformal perturbation theory, we have started from $M$ decoupled Liouville theories on the Mobius strip and added a bulk (marginal) perturbation $e^{\frac{1}{2}(g_x+g_{x+1})}$. To leading order in this bulk coupling and zeroth order in $\de v$ we can incorporate its effect by evaluating it in the reparametrised saddle configuration~\eqref{eqn:reparametrized saddle2}; see~\cite{Maldacena:2018lmt} for a more detailed discussion. Alternatively, our Ansatz for the field configuration in the IR regime~(\ref{eqn:final field configuration for reparam}) gives the same result. We get:
\begin{align}
    \label{eqn: reparametrized chain}
    I[g]=\sum_{x=0}^{M-1}&\left[-\frac{N \delta v\pi}{4p^2}\int_0^{2\pi}d\tau\, \textrm{Sch}[\tan(f_x/2),\tau]-\mathcal{J}_1^2\frac{N}{4p^2}\int d\tau_1 d\tau_2\left[e^{\frac{1}{2}(g_x(\tau_1,\tau_2)+g_{x+1}(\tau_1,\tau_2))}\right]\right]\nonumber\\=\sum_{x=0}^{M-1}&\Biggl[-\frac{N \delta v\pi}{4p^2}\int_0^{2\pi}d\tau\, \textrm{Sch}[\tan(f_x/2),\tau]+\nonumber\\&-\mathcal{J}_1^2 (\pi \delta v)^2\frac{N}{4p^2}\int d\tau_1 d\tau_2\Biggl[  \frac{ \sqrt{f_x'\left(\tau _1\right)} \sqrt{f_x'\left(\tau _2\right)}}{\sqrt{\sin ^2\left(\frac{ f_x\left(\tau _1\right)-f_x\left(\tau
   _2\right)}{2}\right)}}   \frac{ \sqrt{f_{x+1}'\left(\tau _1\right)}\sqrt{ f_{x+1}'\left(\tau _2\right)}}{\sqrt{\sin ^2\left(\frac{ f_{x+1}\left(\tau _1\right)-f_{x+1}\left(\tau
   _2\right)}{2}\right)}} \Biggr]\,.
\end{align}
Note that we are only justified in keeping the second nonlocal term if it is the same order as the first one, since
there are other modes orthogonal to $f$ that we discarded and that would give an $O(N/p^2)$ action, not reduced by an additional factor $\de v\ll 1$.
 Thus, we demand that $\mathcal{J}_1^2(\pi\delta v)^2\ll1$. Note that in the IR regime $\frac{1}{\mathcal{J}_0}\sim\pi\delta v$, so we can write $\mathcal{J}_1^2(\pi\delta v)^2\sim\frac{\mathcal{J}_1^2}{\mathcal{J}_0^2}\sim\alpha$. Our requirement simply becomes $\alpha\ll 1$, with the final form of the action:
\es{eqn: reparametrized chain2}{
    I[f]=-\frac{N}{ 4 p^2}\sum_{x=0}^{M-1}&\Biggl[ (\pi\delta v)\int_0^{2\pi}d\tau\, \textrm{Sch}[\tan(f_x/2),\tau]+\\&+\alpha\int d\tau_1 d\tau_2\Biggl[  \frac{ \sqrt{f_x'\left(\tau _1\right)} \sqrt{f_x'\left(\tau _2\right)}}{\sqrt{\sin ^2\left(\frac{ f_x\left(\tau _1\right)-f_x\left(\tau
   _2\right)}{2}\right)}}   \frac{ \sqrt{f_{x+1}'\left(\tau _1\right)}\sqrt{ f_{x+1}'\left(\tau _2\right)}}{\sqrt{\sin ^2\left(\frac{ f_{x+1}\left(\tau _1\right)-f_{x+1}\left(\tau
   _2\right)}{2}\right)}} \Biggr]\,.
}
We refer to this action as the Schwarzian chain following~\cite{Altland:2019lne}.

\pagebreak

\section*{Acknowledgments}

We thank Gabriel Cuomo, Felix Haehl, Zohar Komargodski, Henry Lin, Alexey Milekhin, G\'abor S\'arosi, Douglas Stanford, and Joaquin Turiaci for useful discussions. We thank Micha Berkooz, Ronny Frumkin, Ohad Mamroud, and Josef Seitz for sharing the draft of~\cite{Berkooz:2024ifu} with us, and for coordinating the submission of our works. MB is supported by the Maryam Mirzakhani Graduate Scholarship. MM is supported in part by the STFC grant ST/X000761/1. 

For the purpose of open access, the authors have applied a CC BY public copyright licence to any Author Accepted Manuscript (AAM) version arising from this submission.

\appendix
\section{More on bulk contributions}
\label{appendix:bulk contributions}
At zeroth order in the expansion of ${\cal L}_{bulk}$,  we get the following bulk contribution:
\begin{equation}
    \label{eqn:action at order zero}
     S_{bulk,0}(\Lambda)=\frac{N}{2p^2}\int_0^{\pi}d\bar{\tau}\int_{\Lambda \delta v}^{2\pi-\Lambda \delta v}d\delta\tau \, \mathcal{L}_{bulk,0}(\delta\tau,\bar{\tau})\,,
\end{equation}
where
    \begin{align}
      \label{eqn:order zero}
       \mathcal{L}_{bulk,0}(\delta\tau,\bar{\tau})= \ &(\cos (\delta \tau )+3) \csc ^2\left(\frac{\delta \tau }{2}\right)\nonumber\\&+\frac{2 f''\left(\bar{\tau}-\frac{\delta \tau
        }{2}\right) \left(f''\left(\frac{\delta \tau }{2}+\bar{\tau}\right)+f'\left(\frac{\delta \tau }{2}+\bar{\tau}\right)^2 \cot
        \left(\frac{1}{2} \left(f\left(\bar{\tau}-\frac{\delta \tau }{2}\right)-f\left(\frac{\delta \tau }{2}+\bar{\tau}\right)\right)\right)\right)}{f'\left(\bar{\tau}-\frac{\delta \tau }{2}\right) f'\left(\frac{\delta \tau }{2}+\bar{\tau}\right)}\nonumber\\&+\frac{f'\left(\bar{\tau}-\frac{\delta \tau }{2}\right)}{f'\left(\frac{\delta \tau }{2}+\bar{\tau}\right)}\biggl(2\frac{  f''\left(\frac{\delta \tau }{2}+\bar{\tau}\right) \sin
        \left(f\left(\bar{\tau}-\frac{\delta \tau }{2}\right)-f\left(\frac{\delta \tau }{2}+\bar{\tau}\right)\right)}{ \left(\cos \left(f\left(\bar{\tau}-\frac{\delta \tau }{2}\right)-f\left(\frac{\delta \tau }{2}+\bar{\tau}\right)\right)-1\right)}\nonumber\\
       &+\frac{f'\left(\frac{\delta\tau }{2}+\bar{\tau}\right)^2 \left(\cos \left(f\left(\bar{\tau}-\frac{\delta \tau }{2}\right)-f\left(\frac{\delta \tau }{2}+\bar{\tau}\right)\right)+3\right)}{ \left(\cos \left(f\left(\bar{\tau}-\frac{\delta \tau }{2}\right)-f\left(\frac{\delta \tau }{2}+\bar{\tau}\right)\right)-1\right)}\biggr)\,.
    \end{align}
We can rewrite this in $\tau_1,\tau_2$ coordinates:
\begin{align}
    \label{eqn:zero order in t1 t2}
     \mathcal{L}_{bulk,0}(\tau_1,\tau_2)=\ &\frac{1}{8} \biggl(\frac{2 f''(\tau_1) \left(f''(\tau_2)+f'(\tau_2)^2 \cot \left(\frac{1}{2} (f(\tau_1)-f(\tau_2))\right)\right)}{f'(\tau_1) f'(\tau_2)}+\nonumber\\&+\frac{2 f'(\tau_1) \left(f''(\tau_2) \sin (f(\tau_1)-f(\tau_2))+f'(\tau_2)^2 (\cos (f(\tau_1)-f(\tau_2))+3)\right)}{f'(\tau_2) (\cos (f(\tau_1)-f(\tau_2))-1)}+\nonumber\\&+4 \csc ^2\left(\frac{\tau_1-\tau_2}{2}\right)-2\biggr)\,,
\end{align}
which can be rewritten as:
\begin{align}
    \label{eqn: zero order as total derivatives}
    \mathcal{L}_{bulk,0}(\tau_1,\tau_2)= \ &\partial_{\tau_2}\left[\frac{f''\left(\tau _1\right) \left(\log \left(f'\left(\tau _2\right){}^2\right)-2 \log \left(\sin ^2\left(\frac{1}{2} \left(f\left(\tau
   _1\right)-f\left(\tau _2\right)\right)\right)\right)\right)}{8 f'\left(\tau _1\right)}\right]+\nonumber\\&+\partial_{\tau_1}\biggl[\frac{1}{4} f'\left(\tau _2\right) \left(f\left(\tau _1\right)+4 \sin \left(\frac{f\left(\tau _1\right)}{2}\right) \csc \left(\frac{1}{2}
   \left(f\left(\tau _1\right)-f\left(\tau _2\right)\right)\right) \csc \left(\frac{f\left(\tau _2\right)}{2}\right)\right)+\nonumber\\&-\frac{f''\left(\tau
   _2\right) \log \left(\sin ^2\left(\frac{1}{2} \left(f\left(\tau _1\right)-f\left(\tau _2\right)\right)\right)\right)}{4 f'\left(\tau _2\right)}\biggr]+\frac{1}{1-\cos \left(\tau _1-\tau _2\right)}-\frac{1}{4}\,.
\end{align}
We can now evaluate the total derivatives at the boundaries of the integration intervals and expand in $\delta v$: after some computations we get (\ref{eqn:order zero relevant contrib}).

The first order contribution is:
\begin{equation}
    \label{eqn:action at order one}
     S_{bulk,1}=\frac{N}{2p^2}\int_0^{\pi}d\bar{\tau}\int_{0}^{2\pi}d\delta\tau\, \mathcal{L}_{bulk,1}(\delta\tau,\bar{\tau})\,,
\end{equation}
with
    \begin{align}
\label{eqn:first order}
    \mathcal{L}_{bulk,1}(\delta\tau,\bar{\tau})=\ & \frac{1}{8} \Biggl(f'\left(\bar{\tau}-\frac{\delta \tau }{2}\right) \nonumber\\&\biggl(2 \left((\pi -\delta \tau ) \cot \left(\frac{\delta \tau
   }{2}\right)+2\right) f'\left(\frac{\delta \tau }{2}+\bar{\tau}\right) \csc ^2\left(\frac{1}{2} \left(f\left(\bar{\tau}-\frac{\delta
   \tau }{2}\right)-f\left(\frac{\delta \tau }{2}+\bar{\tau}\right)\right)\right)+\nonumber\\&-(-\delta \tau +\sin (\delta \tau )+\pi ) \csc
   ^2\left(\frac{\delta \tau }{2}\right) \cot \left(\frac{1}{2} \left(f\left(\bar{\tau}-\frac{\delta \tau }{2}\right)-f\left(\frac{\delta
   \tau }{2}+\bar{\tau}\right)\right)\right)\biggr)\nonumber\\&+\frac{(-\delta \tau +\sin (\delta \tau )+\pi ) \csc ^2\left(\frac{\delta \tau
   }{2}\right) f''\left(\bar{\tau}-\frac{\delta \tau }{2}\right)}{f'\left(\bar{\tau}-\frac{\delta \tau }{2}\right)}+2\frac{ 2
   \cos (\delta \tau )+4 (\pi -\delta \tau ) \cot \left(\frac{\delta \tau }{2}\right)}{\cos (\delta \tau )-1}+\nonumber\\&+2\frac{(-\delta \tau +\sin (\delta \tau )+\pi )
   \left(f''\left(\frac{\delta \tau }{2}+\bar{\tau}\right)+f'\left(\frac{\delta \tau }{2}+\bar{\tau}\right)^2 \cot \left(\frac{1}{2}
   \left(f\left(\bar{\tau}-\frac{\delta \tau }{2}\right)-f\left(\frac{\delta \tau }{2}+\bar{\tau}\right)\right)\right)\right)}{f'\left(\frac{\delta \tau }{2}+\bar{\tau}\right)(\cos (\delta \tau )-1)}+\nonumber\\&+\frac{12}{\cos (\delta \tau )-1}  \Biggr)\,.
\end{align}
As we noted in Sec.~\ref{section:bulk contribution}, there is no need to consider a cutoff, as it would only result in contributions that are subleading in $\delta v$. Rewriting eq (\ref{eqn:first order}) in $\tau_1,\tau_2$ coordinates yields:
\begin{align}
    \label{eqn:first order in t1 t2}
    \mathcal{L}_{bulk,1}(\tau_1,\tau_2)=\ &\frac{1}{8} \biggl(f'\left(\tau _1\right) \biggl(2 \left(\left(\tau _1-\tau _2+\pi \right) \cot \left(\frac{1}{2} \left(\tau _2-\tau
   _1\right)\right)+2\right) f'\left(\tau _2\right) \csc ^2\left(\frac{1}{2} \left(f\left(\tau _1\right)-f\left(\tau
   _2\right)\right)\right)+\nonumber\\&+\left(-\tau _1+\tau _2+\sin \left(\tau _1-\tau _2\right)-\pi \right) \csc ^2\left(\frac{1}{2} \left(\tau _2-\tau
   _1\right)\right) \cot \left(\frac{1}{2} \left(f\left(\tau _1\right)-f\left(\tau _2\right)\right)\right)\biggr)+\nonumber\\&+\frac{\left(\tau _1-\tau
   _2-\sin \left(\tau _1-\tau _2\right)+\pi \right) \csc ^2\left(\frac{1}{2} \left(\tau _2-\tau _1\right)\right) f''\left(\tau
   _1\right)}{f'\left(\tau _1\right)}+\nonumber\\&+2\frac{\left(\tau _1-\tau _2-\sin \left(\tau _1-\tau _2\right)+\pi \right)
   \left(f''\left(\tau _2\right)+f'\left(\tau _2\right){}^2 \cot \left(\frac{1}{2} \left(f\left(\tau _1\right)-f\left(\tau
   _2\right)\right)\right)\right)}{f'\left(\tau _2\right)(\cos \left(\tau _1-\tau _2\right)-1)}+\nonumber\\&+2\frac{ \left(+2 \cos \left(\tau _1-\tau _2\right)+4 \left(\tau _1-\tau _2+\pi \right) \cot
   \left(\frac{1}{2} \left(\tau _2-\tau _1\right)\right)+6\right)}{\cos \left(\tau _1-\tau _2\right)-1}\biggr)\,,
\end{align}
which can be written as
\begin{align}
    \label{eqn:first order as total derivatives}
    \mathcal{L}_{bulk,1}(\tau_1,\tau_2)=\ &\partial_{\tau_2}\left[\frac{\left(\tau _1-\tau _2+\pi \right) \cot \left(\frac{1}{2} \left(\tau _1-\tau _2\right)\right) f''\left(\tau _1\right)}{4 f'\left(\tau
   _1\right)}\right]+\nonumber\\&+\partial_{\tau_1}\biggl[\frac{1}{8 f'\left(\tau _2\right)}\left(\left(\tau _1-2 \tau _2+2 \pi \right) \sin \left(\frac{\tau _1}{2}\right)-\tau _1 \sin \left(\frac{1}{2} \left(\tau _1-2 \tau
   _2\right)\right)\right)\nonumber\\& \csc \left(\frac{1}{2} \left(\tau _1-\tau _2\right)\right) \csc \left(\frac{\tau _2}{2}\right) f''\left(\tau
   _2\right)\biggr]+\nonumber\\&+\partial_{\tau_1}\biggl[\frac{1}{8} \left(\left(\tau _1-2 \tau _2+2 \pi \right) \sin \left(\frac{\tau _1}{2}\right)-\tau _1 \sin \left(\frac{1}{2} \left(\tau _1-2 \tau
   _2\right)\right)\right)\nonumber\\&\, \csc \left(\frac{1}{2} \left(\tau _1-\tau _2\right)\right) \csc \left(\frac{\tau _2}{2}\right) f'\left(\tau _2\right)
   \cot \left(\frac{1}{2} \left(f\left(\tau _1\right)-f\left(\tau _2\right)\right)\right)\biggr]+\nonumber\\&+\partial_{\tau_2}\biggl[\frac{1}{8} \left(\tau _2 \sin \left(\tau _1-\frac{\tau _2}{2}\right)+\left(\tau _2-2 \left(\tau _1+\pi \right)\right) \sin \left(\frac{\tau
   _2}{2}\right)\right)\nonumber\\&\, \csc \left(\frac{\tau _1}{2}\right) \csc \left(\frac{1}{2} \left(\tau _1-\tau _2\right)\right) f'\left(\tau _1\right)
   \cot \left(\frac{1}{2} \left(f\left(\tau _1\right)-f\left(\tau _2\right)\right)\right)\biggr]+\nonumber\\&+\partial_{\tau_1}\biggl[\frac{1}{4} \left(\tau _2-\pi \right) \cot \left(\frac{\tau _2}{2}\right) f'\left(\tau _2\right) \cot \left(\frac{1}{2} \left(f\left(\tau
   _1\right)-f\left(\tau _2\right)\right)\right)\biggr]+\nonumber\\&+\partial_{\tau_2}\biggl[-\frac{1}{4} \left(\left(\tau _1+\pi \right) \cot \left(\frac{\tau _1}{2}\right)-4\right) f'\left(\tau _1\right) \cot \left(\frac{1}{2}
   \left(f\left(\tau _1\right)-f\left(\tau _2\right)\right)\right)\biggr]+\nonumber\\&+\frac{\cos \left(\tau _1-\tau _2\right)-2 \left(\tau _1-\tau _2+\pi \right) \cot \left(\frac{1}{2} \left(\tau _1-\tau _2\right)\right)+3}{2
   \left(\cos \left(\tau _1-\tau _2\right)-1\right)}\,.
\end{align}
As we did for the order zero contribution, we now need to evaluate the total derivatives at the boundaries of the integration intervals and after some further manipulation, we get (\ref{eqn:first order nice}).
\bibliographystyle{JHEP}

\bibliography{biblio}

\end{document}